\documentclass[twocolumn,aps,showpacs,preprintnumbers,superscriptaddress]{revtex4}
\usepackage{graphicx}
 
\usepackage{bm}
\usepackage{amsmath, amsthm, amssymb}

\begin{document}
\title{Linear control of light scattering with  multiple coherent light excitation}
\author{Jeng Yi Lee}
\affiliation{Department of Applied Science, National Taitung University, Taitung 950, Taiwan}

\author{ Yueh-Heng Chung}
\affiliation{
Institute of Photonics Technologies, National Tsing-Hua University, Hsinchu 300, Taiwan}

\author{Andrey E. Miroshnichenko}
\affiliation{School of Engineering and Information Technology, University of New South Wales Canberra, ACT 2600, Australia}

\author{Ray-Kuang Lee}
\affiliation{
Institute of Photonics Technologies, National Tsing-Hua University, Hsinchu 300, Taiwan}
\affiliation{Physics Division, National Center of Theoretical Sciences, Hsinchu 300, Taiwan}
\date{\today}

\begin{abstract}
With the wave interferometric approach, we study how extrinsically multiple  coherent waves excitation can dramatically alter the overall scattering states, resulting in tailoring the energy assignment among radiation and dissipation.
To explore the concept, we derive the corresponding formulas for dissipation and scattering powers for cylindrical passive systems encountered by general configurations of incident waves with various illuminating directions, phases, and intensities.
We demonstrate that a linear superposition of incident waves extrinsically interferes the target channels in a desirable way.
Moreover, the interferometric results can be irrespective to the inherent system configurations like size, materials, and structures.
The extrinsic interfering waves pave a non-invasive solution to manipulate light and matter interaction, with potential applications in metasurfaces, nanophotonics, and metadevices.
\end{abstract}

\maketitle

\section{Introduction}
Waves interference has provided the route to create non-trivial polarization, spin-orbital coupling, Poynting energy and electromagnetic wave distribution.
The interferometric mechanism would lead to systems having peculiar functionalities.
With a perfect choice of phases and intensities in the input waves, coherent perfect absorbers (CPA) enable the out-going waves destructively suppressed without any scattering, as well as to achieve constructively waves inside system, causing totally light trapped \cite{CPA,linear}.
In addition to one dimension system dealing only with forward and backward waves, for high dimensional scattering systems, e.g., finite-sized obstacles, the characteristic of scattering and irradation waves can be decomposed into multiple multipolar series~\cite{book3,book1,multipolar}.
The knowledge of the multipolar is crucial to tailor the radiation pattern, polarization conversion, energy management,  and phase manipulation.
The corresponding scattering coefficients are related to the material parameters, system structures, polarized fields, impinging wave forms, and illumination direction.
When embedded well designed structures and material parameters, a scatterer system could exhibit magnetic meta-atom~\cite{magnetic1,magnetic2,magnetic3}, invisible cloaking~\cite{invisible1,invisible2}, superdirective antenna~\cite{super1}, perfect absorption object~\cite{super2,phase1}, Kerker scatterings\cite{phase2,phase3}, anapole~\cite{anapole}, and superscatterings~\cite{superscattering}.

Instead of single plane wave excitation, recently, it is found that under designed fields, the scattered signature of small particles can reduce the dominant electric dipole,  as well as enhance the weaker magnetic response, resulting in forms of asymmetrical Kerker effects \cite{sensitive1}.
It is also found that the local electromagnetic response of light scattering of extremely small objects  can offer opportunities for locating sub-angstrom dimension in spatial variation of phase and intensity in structured waves~\cite{sensitive2,sensitive3}.
However, it is unclear  how the light scattering and absorption are  manipulated by the linear superposition of various incident waves excitation.

In this work, we derive the formulas of the scattering and absorption powers for cylindrical systems encountered by coherent multiple-incident-waves.
An interfering factor constituted by the linear superposition of excitation waves is revealed.
This factor responds to the manipulation on  the overall light scattering response, which not only modulates the resulting scattering channels, but also affects the scattering distribution and dissipation energy outside and inside the scatterer, respectively.
In particular, we implement a set of impinging coherent waves to turn off the target channels, i.e., both for scattering radiation and dissipation energy.
To achieve the same scattering response, more than one of solutions combined with the other incident waves exist.
Interestingly,  one can open the same scattering channels  with different irradiation configurations.
In addition, the interferometric mechanism can break the rotational symmetry of scattering coefficient $a_n=a_{-n}$, which means that we can suppress the $n$-th channel without affecting the $-n$-th channel.
The results in this work pave a useful non-invasive way to manipulate light-material interaction at nano scales.

\section{Theory of multiple coherent waves excitation}
We consider a scatterer with the cylindrical symmetry, which is illuminated by a  monochromatic TE wave, i.e., the polarized magnetic field is along $z$ axis.
The radius of our cylindrical object is denoted as $a$.
The time evolution of wave is chosen as $e^{-i\omega t}$, with the angular frequency  $\omega$.
The direction of wave vector propagation is on the $x-y$ plane with an angle $\Phi_1$ with respect to $x$ axis.
The corresponding magnetic field of our incident wave can be decomposed into a coherent sum of cylindrical waves, i.e.,  Jacobi-Anger expansion~\cite{book2}:
\begin{equation}\label{incident}
\vec{H}_{in}^{(1)}(\theta,r)=\hat{z}\sum_{n=-\infty}^{\infty}i^{n}e^{in\theta-in\Phi_1}H_1 J_n(k_0r),
\end{equation}
here $J_n$ is Bessel function, $k_0$ is the environmental wavenumber, and $H_1$ is denoted as the complex amplitude of incident magnetic wave.
We note that the propagation direction is associated with the term $e^{-in\Phi_1}$ in this series.
The associated scattered wave can be expressed as
\begin{equation}
\vec{H}_{scat}^{(1)}=\hat{z}\sum_{n=-\infty}^{\infty}i^{n}e^{in\theta-in\Phi_1}H_1 a_n^{TE}H^{(1)}_n(k_0r),
\end{equation}
with the Hankel function of the first kind  $H_n^{(1)}$, representing the out-going wave, and  the complex scattering coefficient $a_n^{TE}$ which is determined by boundary conditions.
 In this convergent series, each term represents a unique scattering modes: $n=0$ for the magnetic dipole (MD), $n=\pm 1$ for the electric dipole (ED), and $n=\pm 2$ for the electric quadrupole (EQ).
 The corresponding electric field can be found by using Maxwell-Amp\'ere equation.
Details for the derivation and formula on the scattering coefficient of TE wave, $a_{n}^{TE}$ are given in Appendix A.
 
Now, for multiple coherent waves excitation,  the resulting incident wave can be obtained based on linear superposition of Eq. (\ref{incident}):
 \begin{equation}
 \begin{split}
\vec{H}_{in}&=\vec{H}_{in}^{(1)}+\vec{H}_{in}^{(2)}+..+\vec{H}_{in}^{(N)}\\
&=\hat{z}\sum_{n=-\infty}^{\infty}i^ne^{in\theta}\sum_{m=1}^{m=N}e^{-in\Phi_m}H_m J_n(k_0r).
\end{split}
 \end{equation}
 Here, we assume that there are $N$ incident waves and each of them has its own propagation direction $\Phi_m$ and the complex amplitude $H_m$.
As a consequence, the corresponding total scattering waves become
  \begin{eqnarray}
\vec{H}_{scat}&=&\vec{H}_{scat}^{(1)}+\vec{H}_{scat}^{(2)}+..+\vec{H}_{scat}^{(N)}\\ \nonumber
&=&\hat{z}\sum_{n=-\infty}^{\infty}i^ne^{in\theta}H^{inf}_n a_n^{TE}H_n^{(1)}(k_0r).
\end{eqnarray}
Here, an interfering factor involving all the excitation is introduced, i.e.,
\begin{eqnarray}
H^{inf}_n=\sum_{m=1}^{m=N}e^{-in\Phi_m}H_m.
\end{eqnarray}
We will illustrate that this interfering factor,  $H_n^{inf}$, plays a crucial role in the interferometric effect to tune the scattering fields both for outside and inside the scatterer.
It is remarked that this interfering term also depends on the channel index $n$, giving different effects on different channels.
 
 To obtain the total scattering and absorption powers,   we apply the Poynting power vectors, as well as  the asymptotic analysis, to calculate the net power integrated over a closed area.
 In Appendix B, one can find the deviations in detail.
 The corresponding absorption and scattering powers for  multiple coherent waves excitation are 
 \begin{eqnarray}
 \label{control}
 P_{abs}^{TE}&=&-\frac{2}{k_0}\sqrt{\frac{\mu_0}{\epsilon_0}}\sum_{n=-\infty}^{n=\infty}\vert H_n^{inf} \vert^2 \{Re[a_n^{TE}]+\vert a_n^{TE}\vert^2\},\nonumber\\
 P_{scat}^{TE}&=&\frac{2}{k_0}\sqrt{\frac{\mu_0}{\epsilon_0}}\sum_{n=-\infty}^{n=\infty}\vert H_n^{inf} \vert^2 \vert a_n^{TE}\vert^2.
 \end{eqnarray}
 Here, $\epsilon_0$ and $\mu_0$ are environmental permittivity and permeability, respectively.
 
 It is known that the scattering coefficient is related to the choices of structure, electromagnetic materials, and operation frequency, which also reflects energy assignment through the scattering (radiative) and  absorption (dissipative) powers.
In most studies, to achieve anomalous scatterers for specific purposes, people always seek for a proper structure design with electromagnetic materials embedded.
However, in this work, we reveal that the interfering factor $H_n^{inf}$ will provide an extrinsic way to control the scattering and field patterns outside and inside the scatterer.

It is noted that the scattering coefficient has a symmetry property when one replaces $n\leftrightarrow -n$.
 However, in the following, we would exhibit the breaking in this symmetry property by imposing a proper set of  multiple incident waves.
In the following, without loss of generality, we consider a single layer cylinder embedded by isotropic and homogeneous electromagnetic material. Nevertheless, our discussion and conclusion can be easily applied to  other structures. 
 
Moreover, instead of TE wave illumination,  one can have formulas for TM waves illumination, i.e., the polarized electric waves along $z$ axis, by replacing, $\epsilon\rightarrow\mu$, $\mu\rightarrow\epsilon$, $\vec{E}\rightarrow \vec{H}$, and $\vec{H}\rightarrow -\vec{E}$, respectively, in view of electromagnetic duality.

\begin{figure*}
\includegraphics[width=17.0cm]{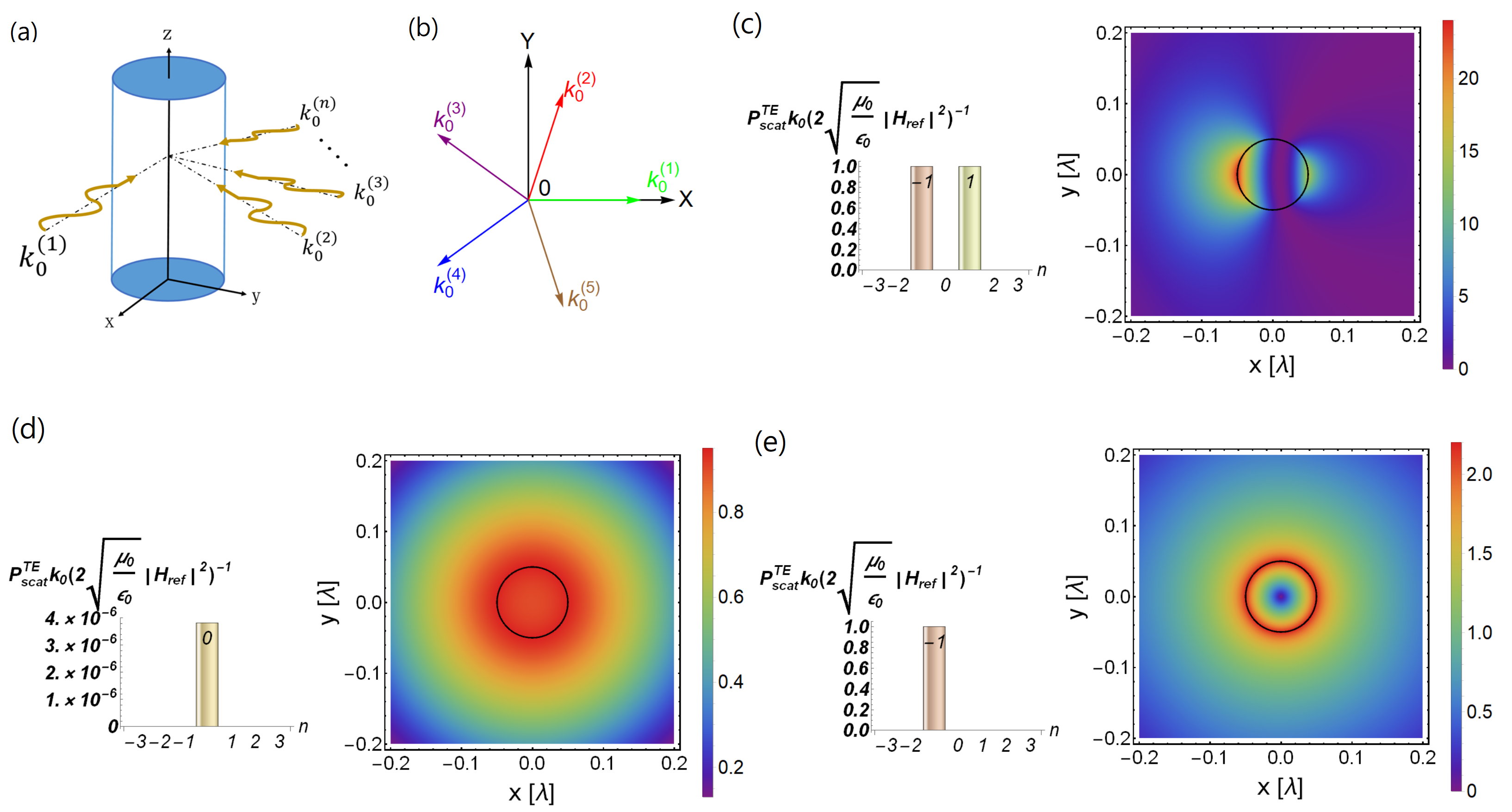}
\caption{(a) Illustration of our scattering system with multiple coherent waves excitation. (b) An example of five incident TE waves on the scatterer. (c) For a single TE wave excitation along $x$-axis.  Left: channels analysis for the normalized scattering powers for our scatterer, revealing the  resonant electric dipoles at $n=-1$-th and $n=1$-th channels. Right:  the resulting intensity of magnetic field.
 (d) For multiple (five) coherent waves excitation.  Left: the normalized scattering powers when irradiated by the (b) scenario.  One can see now,  the $n=0$-th (MD)  channel becomes important, although its strength is quite small. Right: the intensity of  magnetic field, revealing the MD pattern. (e)  With another set of wave amplitudes, but with the same arrangement of illumination directions in (b). Left: the normalized scattering powers in each channels, revealing only  $n=-1$-th (ED) channel survived. Right: the intensity of magnetic field revealing this $n=-1$-th ED pattern.
 Note that, all the field plots are depicted in unit of $\vert H_{ref} \vert^2$. }
\end{figure*}

\section{Linear Control of light scattering and absorption via the interfering factor}
In this section, we demonstrate that a proper setting on the  excitation waves, with the right phases and intensities, can be used to manipulate the light scattering and absorption in a desired way.

\subsection{Multiple TE waves excitation}
Firstly, we consider five incident waves with TE polarization, as illustrated in Fig. 1(a).
The corresponding complex amplitudes are defined as $[H_1, H_2, H_3, H_4, H_5]$,  illuminating from different impinging angles, $[\Phi_1,\Phi_2,\Phi_3,\Phi_4,\Phi_5]$.
The corresponding interfering factor has the form:
$H^{inf}_n= H_1 e^{-in\Phi_1}+H_2 e^{-in\Phi_2}+H_3 e^{-in\Phi_3}+H_4 e^{-in\Phi_4}+H_5 e^{-in\Phi_5}$. 
For each excitation wave,  there are three degrees of freedom: direction of illumination $\Phi_i$, intensity $\vert H_i \vert ^2$, and its phase $\text{Arg}[H_i]$.

Then, we can obtain the following relationship between the interfering factors for the lowest  five channels (from $n=-2$ to $n=2$) and the corresponding five excitation waves in a compact matrix form:
\begin{widetext}
\begin{eqnarray}\label{matrix}
\begin{bmatrix}
\exp[i2\Phi_1] &\exp[i2\Phi_2] &\exp[i2\Phi_3] &\exp[i2\Phi_4] &\exp[i2\Phi_5] \\
\exp[i\Phi_1] &\exp[i\Phi_2] &\exp[i\Phi_3] &\exp[i\Phi_4] &\exp[i\Phi_5] \\
1 & 1 & 1 & 1 & 1\\
\exp[-i\Phi_1] &\exp[-i\Phi_2] &\exp[-i\Phi_3] &\exp[-i\Phi_4] &\exp[-i\Phi_5] \\
\exp[-i2\Phi_1] &\exp[-i2\Phi_2] &\exp[-i2\Phi_3] &\exp[-i2\Phi_4] &\exp[-i2\Phi_5] \\
\end{bmatrix}
\begin{bmatrix}
H_1\\
H_2\\
H_3\\
H_4\\
H_5\\
\end{bmatrix}=
\begin{bmatrix}
H_{-2}^{inf}\\
H_{-1}^{inf}\\
H_{0}^{inf}\\
H_{1}^{inf}\\
H_{2}^{inf}\\
\end{bmatrix}.
\end{eqnarray}
\end{widetext}
The obtained matrix implies that when we assign the desired interfering factors as well as the illumination directions,  the corresponding five incident waves $[H_1, H_2,H_3,H_4,H_5]$ could be found.

Now, we choose the irradiation system with  $[\Phi_1=0,\Phi_2=2\pi/5,\Phi_3=4\pi/5,\Phi_4=6\pi/5,\Phi_5=8\pi/5]$, as depicted in Fig. 1(b).
Then, we consider a lossless scatterer with electric dipole resonance, i.e., $\vert a^{TE}_1\vert =\vert a^{TE}_{-1}\vert=1$ as our studying system.
The material parameters embedded are lossless with the relative permeability $\epsilon_1=-1.156$ and relative permeability $\mu_1=1$. 
As for the cylindrical scatterer, the radius is chosen as $a=0.05\lambda$.
As a comparison, in Fig. 1(c), we show the normalized scattering powers in each channels for a single TE wave excitation along $x$-axis.
As one can see, the Left panel of 1(c), the dominant channels are $a_{-1}^{TE}$ and $a_{1}^{TE}$ ~\cite{ps}.
The corresponding intensity of  magnetic field, reflects the resonant electric dipole, as shown in the Right panel of 1(c).
Here, the magnitude of this single  magnetic wave is defined as $H_{ref}$.
 
\begin{figure*}[th]
\includegraphics[width=18.0cm]{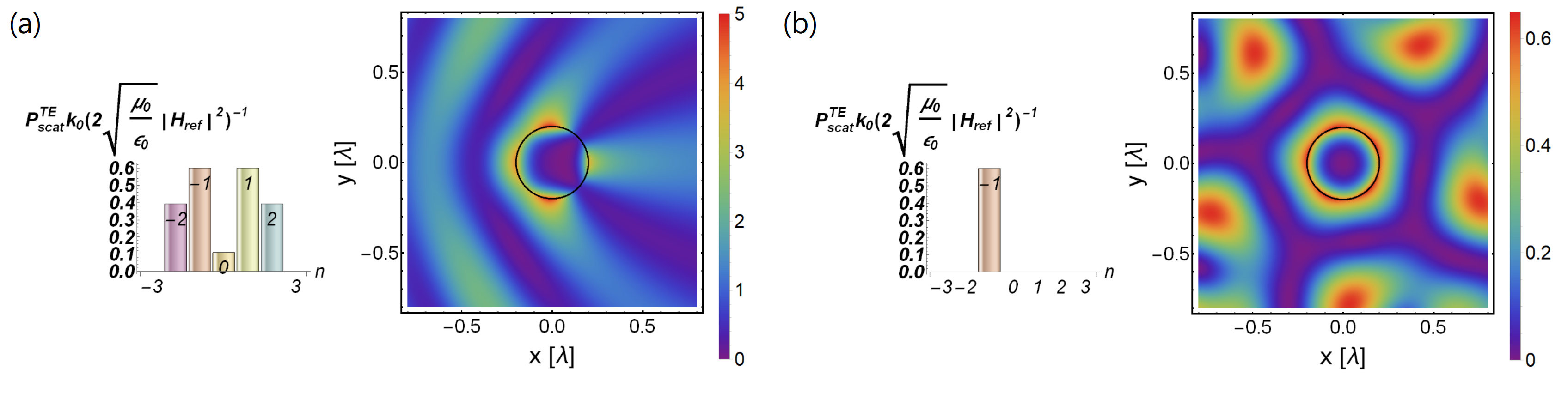}
\caption{Comparison between (a) single and (b) multiple excitations for a large scatterer. (a) Left: the normalized scattering powers in each channel. There are three dominant channels: $n=0$ (MD), $n=\pm 1$ (ED), and $n=\pm 2$ (EQ). Right: the corresponding intensity of magnetic fields. (b) Left: under the same illumination setting, but now the $n=-2$ (EQ), $n=0$ (MD), $n=1$ (ED), and $n=2$ (EQ) are destructively interfered, except for $n=-1$ (ED). Right: the intensity of magnetic field displaying a dipole scattering pattern inside. Again, in all field plots, the values are depicted in unit of $\vert H_{ref} \vert^2$.}
\end{figure*}

Suppose, we want to eliminate all the lowest channels, i.e.,  $n=-2$ (EQ), $n=-1$ (ED), $n=1$ (ED), and $n=2$  (EQ), but keep  $n=0$-th (MD) channel survived.
Then, the extrinsic interfering factors for each channel should be chosen as $[H_{-2}^{inf}=0, H_{-1}^{inf}=0, H_{1}^{inf}=0,H_{2}^{inf}=0]$, and $H_{0}^{inf}=1H_{ref}$.
By solving the matrix equation in Eq.(\ref{matrix}), the corresponding incident wave amplitudes are found: $[H_1=0.2H_{ref}, H_2=0.2H_{ref},H_3=0.2H_{ref},H_4=0.2H_{ref},H_5=0.2H_{ref}]$.

In Fig. 1(d), we shows the corresponding result for the normalized scattering powers in each channel $n$, defined as $P_{scat}^{TE}k_0(2\sqrt{\mu_0/\epsilon_0}\vert H_{ref} \vert^2)^{-1}$.
With the comparison to the scenario in Fig. 1(c), clearly one can observe the expected zeros in EQ and ED,  due to the destructive conditions meet, see Left panel  of Fig. 1(d).
Nevertheless, for the $n=0$-th (MD) channel, it gives a non-zero value.
For the intensity of magnetic field shown in Right panel of Fig. 1(d), we have a MD pattern with $a^{TE}_0$ inside the system.
It demonstrates that via extrinsically interferometric waves, one can alter the scattering states.

To illustrate the flexibility in controlling light scattering by multiple coherent waves excitation. 
In Fig. 1(e), we demonstrate another setting for the interfering factors by choosing $[H_{-2}^{inf}=0,H_{-1}^{inf}=1H_{ref},H_{0}^{inf}=0,H_{1}^{inf}=0,H_{2}^{inf}=0]$, as well as the same illumination directions in Fig. 1 (b).
This set represents the destructive interferences at $n=-2$, $n=0$, $n=1$, and $n=2$-th channels, but constructive interference at the $n=-1$-th channel.
The corresponding incident wave amplitudes are calculated as $[H_{-2}=0.2H_{ref}, H_{-1}=(0.06-0.2i)H_{ref}, H_{0}=(-0.16-0.12i)H_{ref}, H_{1}=(-0.16+0.12i)H_{ref}, H_{2}=(0.06+0.2i)H_{ref}]$.

Now, we find that these incident amplitudes become complex, due to the choice of asymmetrical interfering factors in channels.
The corresponding normalized scattering powers, shown in Fig. 1(e), support our goal, i.e., only a non-zero value happened at the $n=-1$-th channel.
The intensity plot of magnetic field is shown in Right panel of Fig. 1(e), displaying the electric dipole preserved in $\vert a^{TE}_{-1}\vert =1$.
Even though the rotational symmetry of this cylindrical scatterer leads to the complex scattering coefficient possessing $a_{n}=a_{-n}$, here,  we break this symmetry by using the extrinsic interferometric mechanism.

 \begin{figure*}[t]
\includegraphics[width=18.0cm]{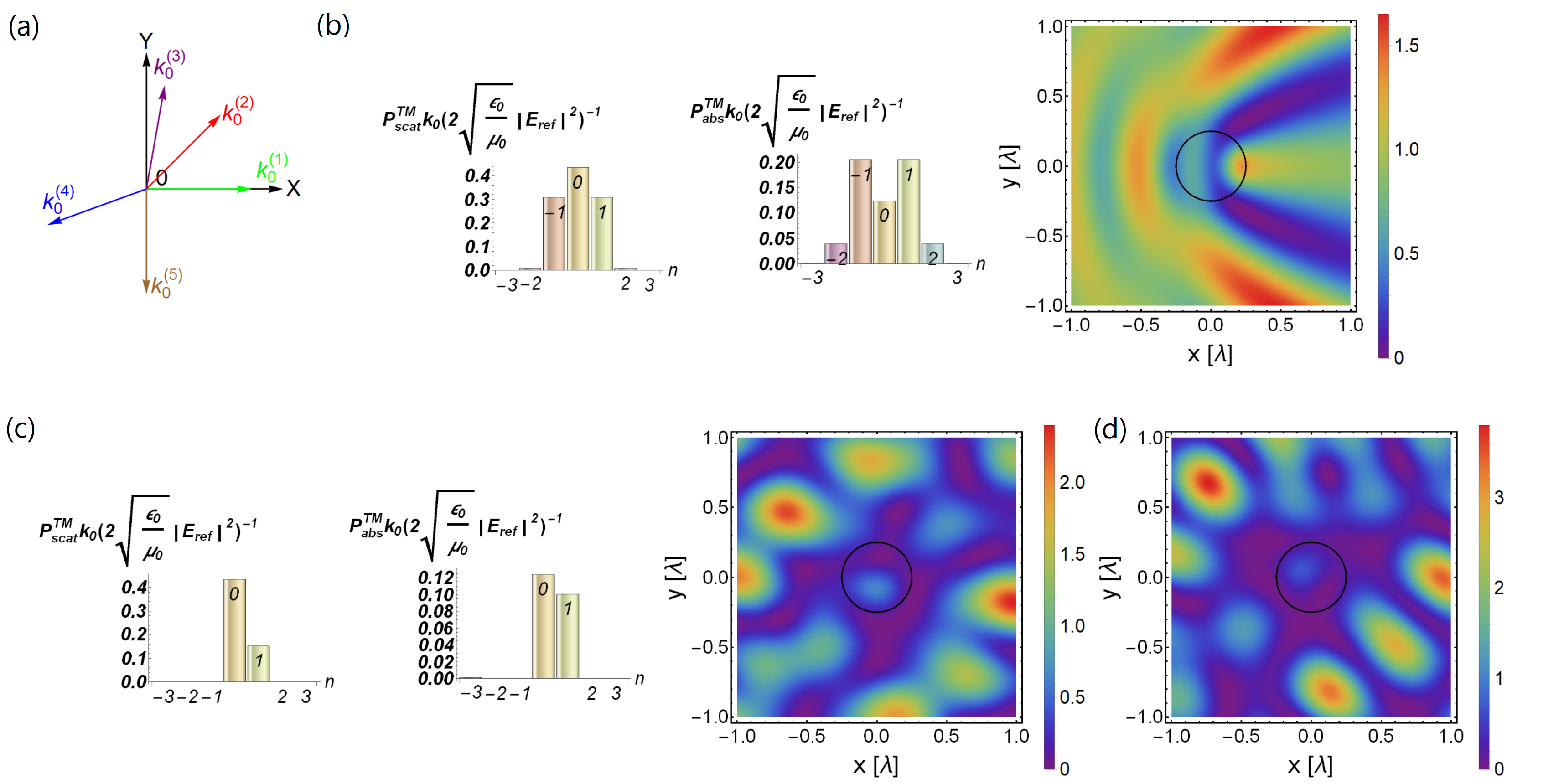}
\caption{(a) Illustration of our scattering system with five TM wave excitations. (b) For a single TM wave excitation: Left: the normalized scattering and absorption powers. Right: the corresponding intensity of electric field in unit of $\vert E_{ref}\vert^2$. (c) For multiple (five) coherent wave excitations. Left: the normalized scattering and  absorption powers. Right: the corresponding intensity of electric field. (d) Intensity of electric field by another set of complex interfering factors, but their modulus are same as (c).
It is shown that the field around and inside scatterer are different to that of (c), although they all possess same scattering and absorption powers.}
\end{figure*}

\subsection{Size effects}
In order to investigate the size effect on scattering, we choose a larger scatterer with the radius  $a=0.2\lambda$,  compared to $a_0 = 0.05 \lambda$ used in Section III, embedded with a lossless non-magnetic material $\epsilon_1=-3$. 
In Fig. 2(a), we start a single excitation. 
The corresponding intensity of magnetic field shows some maximum strengthes around the scatterer.
Instead, by utilizing the same multiple illumination setting given in Fig. 1(b), along with the same choice of incident wave amplitudes in Fig. 1(e), only $a^{TE}_{-1}$  survives, as shown in Left panel of Fig. 2(b).
With a different size in the scatterer, even though the resulting scattering components in each channels are different to those shown in Fig. 1(c),  the outcome can be remained for the $n=-1$-th channel.
As one can see in Right panel of Fig. 2(b),   similar ED pattern can be found inside the scatterer as that revealed in Fig. 1(e).
With these results, we can say that  the interferometric effect is beyond any geometry size, as the interfering factor filters and modulates the target channels.

\subsection{Multiple TM wave excitation}
In addition to TE wave excitation, we can apply our concept to TM wave excitations as well.
Here, we also choose  five TM waves excitation, i.e., the electric waves are polarized along $z$-axis.
The illumination directions are chosen as $[\Phi_1=0,\Phi_1=0.785398,\Phi_1=1.39626,\Phi_1=3.49066,\Phi_1=4.71239]$, as illustrated in  Fig. 3 (a).
Here, we let the interfering factors in five channels be $[E_{-2}^{inf}=0,E_{-1}^{inf}=0,E_{0}^{inf}=1E_{ref}, E_{1}^{inf}=0.7E_{ref},E_{2}^{inf}=0]$, in order to  create fully destructive conditions in  $n=-2$ (magnetic quadrupole MQ), $n=-1$ (MD), and $n=2$ (MQ) channels, but constructive conditions in $n=0$ (ED) and $n=1$ (MD) channels.
Unlike the scenarios for TE wave excitations, as shown in Figs. 1 and 2, this set of interfering factors gives different weightings in the $n=0$ and $n=1$ channels. 
The corresponding wave amplitudes can be found as: $[E_{-2}=(0.5+0.028i)E_{ref},E_{-1}=(-0.19-0.09i)E_{ref},
E_0=(0.38+0.23i)E_{ref},
E_1=(0.09+0.01i)E_{ref},
E_2=(0.212-0.18i)E_{ref}]$, with the reference wave amplitude $E_{ref}$.

As for the scatterer, we choose a lossy non-magnetic  material, with $\epsilon_1=2.3+0.5i$  and the radius $a=0.25\lambda$.
Under a  single wave excitation, in Fig. 3(b), we show the corresponding  normalized absorption (defined as $P_{abs}^{TM}k_0(2\sqrt{\epsilon_0/\mu_0}\vert E_{ref} \vert^2)^{-1}$) and scattering powers (defined as $P_{scat}^{TM}k_0(2\sqrt{\epsilon_0/\mu_0}\vert E_{ref} \vert^2)^{-1}$ ).
One can see from Left panel of Fig. 3(b), there are three scattering channels dominant, and the  corresponding intensity of electric field  is also depicted in  Right panel of Fig. 3(b).

Now, for the multiple TM wave excitation, with the same arrangement of incident waves in Fig. 3(a),  only two channels, $n=0$ and $n=1$,  remain non-zero in the resulting scattering and absorptive powers, see Left panel of Fig. 3(c).
With the comparison to the single wave excitation, we can see that the absorption power in the $n=0$-th channel is lower than that in $n=1$-th channel.
However, with the multiple illumination, the strengths of normalized absorption powers in the $n=0$-th channel becomes larger than that in the $n=1$-th channel.
 As the lossy scatterer is considered here, our result indicates that the interferometric technology can also  tune the dissipative energy assignments in a designed ways.
 
In addition, in all the examples illustrated above, we have limited ourself to have the interfering factors being real.
However, we can release this constraint by using a complex value while keeping the modulus of interfering factors fixed.
Now, we can choose the interfering factors as: $[E_{-2}^{inf}=0,E_{-1}^{inf}=0,E_{0}^{inf}=1E_{ref}, E_{1}^{inf}=0.7(\cos(0.7\pi)+i\sin(0.7\pi))E_{ref},E_{2}^{inf}=0]$, with the corresponding incident waves having the amplitudes: $[E_1=(0.34+0.05i)E_{ref},E_2=(-0.373+0.178i)E_{ref},E_3=(0.38-0.23i)E_{ref},E_4=(0.38-0.15i)E_{ref},E_5=(0.27+0.15i)E_{ref}]$.
 As shown in Fig. 3(d), the resulting fields are totally different to those shown in Fig. 3(c).
 Nevertheless, the corresponding scattering and absorption powers are all the same.
 Based on these results, the powerful interferometric technology  provides a flexible degree of freedom to achieve the same scattering and dissipation powers, but with to have a variety of irradiation combinations. 

Before conclusion, we emphasize that the setting of irradiative waves is not unique.
As implied by Eq. (\ref{matrix}), we can always find solutions with various irradiation angles and amplitudes to achieve the same interfering results.
Moreover, it is worth to mention that from the intensity plots, one can see non-trivially electric or magnetic field distribution inside the scatterers, as  the scattering response in our system is fully solved by the exact Mie theory, beyond electrically small approximation.
Even though, in this work, we illustrate the concept for five channels with  five wave excitations, the methodology can be easily generalized to deal with more scattering channels.
When the number of excitations is the same as that of controlled channels,  mathematically it is linearly solvable~\cite{ps2}.
Moreover, in the case of CPA, one needs to know the size geometry and electromagnetic materials precisely, in order to realize these functionalities.
However,  in our proposed system, the functionalities are independent of these elements, but dependent on incident waves irradiation.
Least but not last, with a proper set of multipole series~\cite{multipolar},
the methodology developed in this work can also be extended to other finite-sized objects other than cylinders.




\section{Conclusion}
In summary, based on the interferometric concept, we study the scattering in cylindrical systems upon multiple coherent waves excitations.
We rigorously derive the scattering and dissipation powers for a cylindrical object encountered by multiple TE and TM wave excitations, with arbitrary irradiation angles, amplitudes, and phases.
An interfering factor is introduced  to tailor the overall incident, scattering, and internal fields.
By manipulating the  interfering factors, we not only extrinsically control the scattering characteristics, but also tune the related dissipation loss.
More interestingly,  with a proper design of waves excitations in illumination directions, phases, and intensities,
our result is independent from the system configurations, such as its size, structure, and composed material.

We demonstrate how to have systems having different materials and sizes, but supporting  the same scattering channels.
In addition, we also create destructive interference to break the inherently rotational symmetry of scattering coefficients.
In general, the solution to support designed scattering channels is not unique, which provides a flexible degree of freedom in this interferometric technology.
Our methodology can be easily extended to a more complicated geometry and structure, which offers the route for non-invasion applications in  nano-photonics and meta-device.

\section{Acknowledgement}

This work was supported by Ministry of Science and Technology, Taiwan (MOST) ($107$-
$2112$-M-$143$-$001$-MY3).

\appendix
\renewcommand\thesection{}
\setcounter{equation}{0}
\section*{APPENDIX} \label{appA}

\section{A}
In this Appendix, we derive the scattering coefficient $a_n^{TE}$ for a cylindrical object under a TE wave illuminated with angle $\Phi_1$ to x axis.
Outside the object, the total TE magnetic waves are sum of incident and scattering, so we have
\begin{equation}
\begin{split}
&\vec{H}_{in}^{(1)}(r,\theta)+\vec{H}_{scat}^{(1)}(r,\theta)=\\
&\hat{z}\sum_{n=-\infty}^{\infty}i^{n}e^{in\theta-in\Phi_1}H_{1}[J_n(k_0r)
+a_n^{TE}H_n^{(1)}(k_0r)],
\end{split}
\end{equation}
with the corresponding electric wave calculated by Maxwell-Amp\'ere equation,
\begin{equation}
\begin{split}
&\vec{E}_{in}^{(1)}(r,\theta)+\vec{E}_{scat}^{(1)}(r,\theta)=\\
&-\frac{1}{i\omega \epsilon_0}\{\frac{\hat{r}}{r}\sum_{n=-\infty}^{\infty}i^{n+1}n e^{in\theta}H_1[J_n(k_0r)+a_n^{TE}H_n^{(1)}]-\\
&\hat{\theta}\sum_{n=-\infty}^{\infty} i^{n}e^{in\theta}  H_1 k_0 [J^{'}(k_0r)  + a_n^{TE} H_n^{(1)'}(k_0 r)]\}.
\end{split}
\end{equation}
Here $J^{'}_n(x)\equiv dJ_n(x)/dx$ and $H^{(1)'}_n(x)\equiv dH^{(1)}_n(x)/dx$.
Inside the object, the internal fields $\vec{H}_{int}^{(1)}(r,\theta)$ and $\vec{E}_{int}^{(1)}(r,\theta)$ could be expressed as
\begin{equation}
\vec{H}_{int}^{(1)}(r,\theta)=\hat{z}\sum_{n=-\infty}^{\infty}i^{n}e^{in\theta-in\Phi_1}H_{1}b_n^{TE}J_n(k_1 r)
\end{equation}
and 
\begin{equation}
\begin{split}
&\vec{E}_{int}^{(1)}(r,\theta)=
-\frac{1}{i\omega \epsilon_1}\{\frac{\hat{r}}{r}\sum_{n=-\infty}^{\infty}i^{n+1}n e^{in\theta}H_1 b_n^{TE}J_n(k_1r)\\
&-\hat{\theta}\sum_{n=-\infty}^{\infty}i^{n}e^{in\theta}H_1k_1 b_n^{TE}J^{'}(k_1r)\}
\end{split}
\end{equation}
where $\epsilon_1$ is relative permeability, $k_1$ is the corresponding wavenumber, and $b_n^{TE}$ is complex internal field coefficient.

Applying the continuity of electric and magnetic waves along the cylindrical surface at $r=a$, we obtain the relationship for $a_n^{TE}$ and $b_n^{TE}$,
\begin{equation}
\begin{split}
\begin{bmatrix}
H_{n}^{(1)}(k_0a) & -J_{n}(k_1a)\\
\frac{k_0}{\epsilon_0} H_{n}^{(1)'}(k_0a) & -\frac{k_1}{\epsilon_1}J^{'}_{n}(k_1a)\\
\end{bmatrix}
\begin{bmatrix}
a_n^{TE}\\
b_{n}^{TE}
\end{bmatrix}=\\
\begin{bmatrix}
-J_{n}(k_0a)\\
-\frac{k_0}{\epsilon_0}J_n^{'}(k_0a)
\end{bmatrix}.
\end{split}
\end{equation}

For TM waves excitation, due to electromagnetic duality, the related formulas can be obtained by replacement of $\epsilon\rightarrow\mu$, $\mu\rightarrow\epsilon$, $\vec{E}\rightarrow \vec{H}$, and $\vec{H}\rightarrow -\vec{E}$.
\\

\section{B}
In this appendix, we derive the scattering and absorption powers by multi-coherent-TE-waves.
Applying asymptotic analysis ($r\rightarrow\infty$) for these special functions, the leading terms are:
\begin{equation}
\begin{split}
J_{n}(x)&\approx \sqrt{\frac{2}{\pi x}}\frac{1}{2}[e^{i(x-\frac{n\pi}{2}-\frac{\pi}{4})}+e^{-i(x-\frac{n\pi}{2}-\frac{\pi}{4})}]\\
J_{n}^{'}(x)&\approx \sqrt{\frac{2}{\pi x}}\frac{i}{2}[e^{i(x-\frac{n\pi}{2}-\frac{\pi}{4})}-e^{-i(x-\frac{n\pi}{2}-\frac{\pi}{4})}]\\
H_{n}^{(1)}(x)&\approx \sqrt{\frac{2}{\pi x}}e^{i(x-\frac{n\pi}{2}-\frac{\pi}{4})}\\
H_{n}^{(1)'}(x)&\approx i \sqrt{\frac{2}{\pi x}}e^{i(x-\frac{n\pi}{2}-\frac{\pi}{4})}.\\
\end{split}
\end{equation}

Moreover, the total TE magnetic field at far field regime would be
\begin{equation}
\begin{split}
\vec{H}_{tot}&=\vec{H}_{in}+\vec{H}_{scat}\\
&= \hat{z}\sum_{n=-\infty}^{\infty}i^{n}e^{in\theta}H_{n}^{inf}\{J_n(k_0r)+a_n^{TE}H_n^{(1)}(k_0r)\}\\
&\approx\hat{z}\sum_{n=-\infty}^{\infty} i^{n}e^{i n\theta}\sqrt{\frac{2}{\pi k_0r}}H_{n}^{inf}[\frac{1}{2}e^{i(k_0r-\frac{n\pi}{2}-\frac{\pi}{4})}\\
&+\frac{1}{2}e^{-i(k_0r-\frac{n\pi}{2}-\frac{\pi}{4})}+a_n^{TE}e^{i(k_0r-\frac{n\pi}{2}-\frac{\pi}{4})}].
\end{split}
\end{equation}
For total TE electric field, we take only $\theta$ direction term, because only this term would make a contribution to power transport (along radial component) at far region.
\begin{equation}
\begin{split}
&\vec{E}_{tot}=\vec{E}_{in}+\vec{E}_{scat}\\&\rightarrow \frac{1}{i\omega \epsilon_0}\hat{\theta}\sum_n i^{n}e^{in\theta}k_0\sqrt{\frac{2}{\pi k_0r}}H_{n}^{inf}[\frac{i}{2}e^{i(k_0r-\frac{n\pi}{2}-\frac{\pi}{4})}\\
&-\frac{i}{2}e^{-i(k_0r-\frac{n\pi}{2}-\frac{\pi}{4})}+ia_n^{TE}e^{i(k_0r-\frac{n\pi}{2}-\frac{\pi}{4})}]
\end{split}
\end{equation}

The time-averaged Poynting power is  $\vec{P}_{tot}=\frac{1}{2}Re[\vec{E}_{tot}\times\vec{H}_{tot}^{*}]$.
By using the integration of this power over a closed surface, we obtain the absorption power
\begin{equation}
\begin{split}
&P_{abs}=-
\oint\vec{P}_{tot}\cdot\hat{r}rd\theta\\
&=-2k_0\sqrt{\frac{\mu_0}{\epsilon_0}}\sum_{n=-\infty}^{\infty}\vert H_n^{inf}\vert^2[Re[a_n^{TE}]+\vert a_n^{TE}\vert^2].
\end{split}
\end{equation}
Here the minus in this integration means when the system has energy dissipated, its value would be positive.

Following the same derivation approach, we can
derive the scattering power 
\begin{equation}
P_{scat}=\oint\vec{P}_{scat}\cdot\hat{r}rd\theta=\frac{2}{k_0}\sqrt{\frac{\mu_0}{\epsilon_0}}\sum_{n=-\infty}^{\infty}\vert H_n^{inf}\vert^2 \vert a_n^{TE}\vert^2
\end{equation}
here $\vec{P}_{scat}=\frac{1}{2}Re[\vec{E}_{scat}\times\vec{H}_{scat}^{*}]$ and the scattering power also possesses positive property for any observable objects.
For TM waves, the relevant formulas can be found by  $\epsilon\rightarrow\mu$ and $\mu\rightarrow\epsilon$.


\begin{thebibliography}{99}
\bibitem{ps}
Although this system has the electric dipole resonant, there still have other extreme small residual strengths in other channels.
In particular, we find $\vert a_0^{TE}\vert=0.002$.
\bibitem{ps2}
In all the examples above, we always take a non-zero interfering factor in some channels, in order to protect the emergence of zeros outcome of incident waves.
\end{thebibliography}
\end{document}